\documentclass{aa}  

\usepackage[normalem]{ulem}
\usepackage{graphicx}
\usepackage{txfonts}
\usepackage[dvipsnames]{xcolor}
\usepackage[toc,page]{appendix}
\usepackage{floatrow}
\usepackage{floatrow}
\usepackage{soul}
\newcommand{\teff}{$T_{\mathrm{eff}}$}
\newcommand{\msun}{M$_\odot$}
\newcommand{\logg}{$\log g$}

\newcommand{\shree}[1]{\textcolor{black}{#1}} 
\newcommand{\SG}[1]{\textcolor{black}{#1}} 
\newcommand{\forreferee}[1]{\textcolor{black}{#1}}

\begin{document}

   \title{Observational evidence of third dredge-up occurrence in  S-type stars with initial masses around 1~\msun~\thanks{Based on observations made with the Mercator Telescope, operated on the island of La Palma by the Flemish Community, at the Spanish Observatorio del Roque de los Muchachos of the Instituto de Astrofísica de Canarias. }}

   \author{S. Shetye
          \inst{1,2}
          \and
          S. Goriely\inst{1} \and    
 L. Siess\inst{1} \and
 S. Van Eck\inst{1} \and
          A. Jorissen\inst{1}   \and
          H. Van Winckel\inst{2}    
          }

   \institute{ Institute of Astronomy and Astrophysics (IAA), Université libre de Bruxelles (ULB)
,
              CP  226,  Boulevard  du  Triomphe,  B-1050  Bruxelles, Belgium\\
              \email{Shreeya.Shetye@ulb.ac.be}
         \and
             Institute of Astronomy, KU Leuven, Celestijnenlaan 200D, B-3001 Leuven, Belgium
             }

   \date{Received; accepted }

 
  \abstract
   {S stars are late-type giants with spectra showing characteristic molecular bands of ZrO in addition to the TiO bands typical of M stars. Their overabundance pattern shows the signature of s-process nucleosynthesis. 
Intrinsic, technetium (Tc)-rich S stars are  the first objects, on the Asymptotic Giant Branch (AGB), to undergo third dredge-up (TDU) events.
Gaia exquisite parallaxes now allow to precisely locate these stars in the Hertzsprung-Russell (HR) diagram. Here we report on a population of low-mass, Tc-rich S stars, previously unaccounted for by stellar evolution models. 
}
   {Our aim is to derive parameters of a sample of low-mass Tc-rich S stars and then, by comparing their location in the HR diagram with stellar evolution tracks, to derive their masses and to compare their measured s-process  abundance profiles with recently derived STAREVOL nucleosynthetic predictions for low-mass AGB stars.}
   {The stellar parameters were obtained using a combination of HERMES high-resolution spectra, accurate Gaia Data Release~2~(Gaia-DR2)  parallaxes,  stellar-evolution  models and  newly-designed  MARCS  model  atmospheres  for  S-type  stars.} 
   {
   We report on 6 Tc-rich S stars lying close to the 1~\msun~ (\forreferee{initial mass}) tracks of AGB stars of the corresponding metallicity and above the predicted onset of TDU, as expected. This provides direct evidence for TDUs occurring in AGB stars with \forreferee{initial} masses as low as $\sim$1~\msun~ and at low luminosity, i.e. at the start of the thermally-pulsing AGB. We present AGB models producing TDU in those stars
   \forreferee{with [Fe/H] in the range $-0.25$ to $-0.5$} There is a reasonable agreement between the measured and predicted s-process abundance profiles. For 2 objects however (CD -29$^\circ$5912 and BD +34$^\circ$1698), the predicted C/O ratio and s-process enhancements do not match simultaneously the measured ones.}
   {}


   \maketitle
%

\section{Introduction}

S-type stars are characterized by the presence of ZrO molecular bands on top of the
TiO bands typical of M-type giants \citep{Merrill}, along with a carbon/oxygen (C/O) ratio in between the ones of M-type stars ($\sim$0.5) and those of carbon stars ($>$1).
S-process elements, produced during the Asymptotic Giant Branch (AGB) phase, are overabundant in their atmospheres 
 \citep{smithlambertfeb1990}. 
 The fact that technetium (Tc), an s-process element with no stable isotopes, is detected in some but not all S stars, remained a puzzle until it was discovered that Tc-rich S stars are thermally-pulsing AGB (TP-AGB) stars undergoing  s-process nucleosynthesis and third dredge-up (TDU) events, while Tc-poor S stars are less evolved stars owing their s-process enrichment to a pollution from a former AGB companion which has now turned into an undetected white dwarf \citep{jorissen1988,smith1988,jorissen1993, VanEck-1999,VanEck-2000b, VanEck-2000a}. Enough time elapsed since the mass transfer for Tc to decay completely. Hence, S stars can be divided into two groups, the intrinsic, Tc-rich ones and the extrinsic, binary ones with no Tc. S stars serve as important probes to understand the AGB nucleosynthesis, since  intrinsic S stars are the first stars on the AGB to show signatures of TDU.

An intriguing challenge in our understanding of AGB nucleosynthesis and TDU episodes is the growing series of observations pointing at dredge-up occurring in low-mass stars (<1.5 \msun). 
Such evidences are now more solid thanks to Gaia parallaxes \citep{gdr2} allowing to precisely locate objects in the Hertzsprung-Russell diagram  \citep[HR diagram; e.g., for S-type stars,][S18 hereafter]{shreeya1}.
V915$\>$Aql is an example of such a low-mass (\forreferee{initial mass }$\sim$1~\msun) intrinsic S star from the sample of S18 exhibiting s-process element enrichment (including Tc).

There were previous mentions of low-mass stars experiencing dredge-up. For example  
\cite{vanaarle}  and \cite{2015kenneth} reported strongly s-process-enriched and moderately metal-poor low-luminosity post-AGB stars as an evidence of TDU in stars of low initial masses ($\sim$1 \msun).

\forreferee{On the contrary, standard stellar-evolution models at solar  metallicity produce TDUs only in stars with initial masses larger than 1.4~--~1.5~\msun{} \citep{straniero,bisterzo,cristallo,karakas}, with the exception of \cite{weiss2009} which found TDU already at 1~\msun\ for $Z = 0.02$, but allowing for some overshooting below the convective pulse. This minimum initial mass for the occurrence of TDU actually depends on the chemical composition, mass-loss rate, mixing prescriptions and numerics. 
For example, the models of \cite{lugaro2012} and \cite{fishlock2014} make TDU at $\sim$1~\msun\ but for metallicities as low as [Fe/H]~$\sim -2.3$ and $-1.2$, respectively. Similarly, the AGB models of \cite{stancliffe} and \cite{weiss2009} give rise to TDU at 1~\msun\ for the LMC and SMC metallicities (respectively $Z = 0.008$ and 0.004, as adopted for these computations, which for $Z_\odot = 0.02$ translates to [Fe/H]$_{\rm LMC} = -0.4$ and [Fe/H]$_{\rm SMC} = -0.7$). 
We moreover stress that \cite{stancliffe} did not consider any mass loss at any stage of the evolution.}


\forreferee{Here, we report on the analyses of 6 Tc-rich S stars in the solar neighbourhood. Thanks to the Gaia DR2 parallaxes, we compare the location of the sample stars in the HR diagram with the new AGB models computed using the STAREVOL code to determine their evolutionary masses. We further discuss the agreement between the measured s-process abundances and the nucleosynthesis predictions.}


\section{Observations}

Among the  Tc-rich S stars with an accurate Gaia DR2 parallax (i.e., matching the condition {$\sigma_{\varpi} / \varpi$}~$\leq 0.3$) that we observed with the HERMES high-resolution  spectrograph \citep{raskin}, we selected the stars that appeared to have \forreferee{an initial} mass around 1~\msun. Our sample includes 6 low-mass Tc-rich S stars; V915~Aql, which was analysed in S18, is among them.
Only spectra with a S/N ratio larger than or equal to 30 in the $V$ band were used, to ensure Tc-line detectability. The observation log can be found in Table~\ref{basic data}.  
Stellar masses 
were obtained from the position of the stars in the HR diagram (with luminosities obtained from the Gaia DR2 parallaxes and stellar parameters as derived in Sect.~\ref{sectionparams}) compared with the STAREVOL evolutionary tracks (see also S18 for more details about the way masses were derived).
The parallaxes and other basic data of the sample stars can be found in Table~\ref{basic data}. 


\section{Technetium detection}
The radio-isotope pairs Tc-Mo and  Zr-Nb give a precise diagnostic to decide whether a star is currently experiencing s-process nucleosynthesis and dredge-ups \citep{Mathews1986,pieter2015}.
Intrinsic S stars can thus be identified without ambiguity if they are enriched in Tc \citep{merrill52} but not in Nb \citep{pieter2015,drisya}.
We use the three strong \ion{Tc}{I} resonance lines  located at $4238.19$~\AA, $4262.27$~\AA,~and $4297.06$~\AA.
Though these lines are heavily blended with other (s-process) lines, their combination can be used reliably for the intrinsic/extrinsic classification. 
Fig.~\ref{Tc} displays the absorption features produced by the three lines for our sample stars and illustrates that they are all intrinsic S stars. 
\cite{wang} classified BD~+34$^\circ$1698 and HD~357941 as extrinsic S stars based on their location in the  $(K-[12], [12]-[25])$ color-index plane (where [12] and [25] are the IRAS magnitudes); however, our high-resolution spectra demonstrate without ambiguity that they are Tc-rich. 
The classification of CD~$-29^\circ5912$ as an intrinsic S star is in agreement with the classification by \cite{sophieMarcs}.
There was no former classification available in the literature for CSS$\>$182 and CSS$\>$154.

The second diagnostic, based on the Nb abundance, is in perfect agreement with the Tc diagnostic, as will be shown in Sect.~\ref{abundsection}.


\section{Stellar parameter determination}\label{sectionparams}
The atmospheric parameters \{T$_{\rm eff}$, \logg, [Fe/H], C/O, [s/Fe]\}, where  [s/Fe]\footnote{The abundance of element X is defined as 
$\mathrm {[X/Y]} = \log_{10} (n_{\rm X}/n_{\rm Y})_*- \log_{10}(n_{\rm X}/n_{\rm Y})_\odot$ where $n_i$ is the number density of element $i$, and Y is a normalising element (usually H or Fe). \forreferee{The exact meaning of [s/Fe] in the grid of S-star MARCS model atmospheres is described in Appendix A.1.}} is the s-process enhancement with respect 
to the (metallicity-scaled) solar s-process contribution,
were derived as
in S18. In summary, this method performs a $\chi^2$ adjustment 
between a grid of S-star MARCS synthetic spectra \citep{sophieMarcs}
and observed HERMES spectra within carefully selected spectral regions, also considering synthetic and observed photometric color indices.
\shree{Luminosity was calculated using the distances derived from the Gaia DR2 parallaxes, the reddening $E_{B-V}$ (retrieved from \citealt{gontcharov}) and the bolometric correction in the $K$-band as computed from the MARCS model atmospheres.}
The luminosity combined with $T_{\rm eff}$ leads to a constrain on the stellar mass by comparing with STAREVOL evolutionary tracks, allowing to re-evaluate the surface gravity. \forreferee{The parameter selection  was iterated till the surface gravity derived from the spectroscopic adjustment matched the value derived from the stellar position in the HR diagram}.  
The variations of the atmospheric parameters while iterating for \logg\ are used as an estimate of the parameter uncertainty, as listed in Table~\ref{params}.


\section{Abundance determination and uncertainties}\label{abundsection}
The atomic abundances were derived by comparing observed and synthetic spectra generated by the Turbospectrum (\citealp{rodrigo}, \citealp{turbospectrum}) spectral synthesis code on MARCS model atmospheres of S-type stars \citep{sophieMarcs} with matching parameters. For V915~Aql, the stellar parameters and abundances from S18 were used.\\
\textit{C, N, O}: 
The C/O ratio is obtained from the stellar parameter determination (Sect.~\ref{sectionparams}).
The oxygen abundance cannot be derived from the optical spectrum in S-type stars, hence its solar-scaled value at the stellar metallicity was adopted, thereby fixing the C abundance.
The nitrogen abundance was derived from the CN lines in the 7900-8100~\AA~range. In particular, the lines listed in \cite{tibaulth} were used.\\
\textit{[Fe/H]}: 
The metallicity was derived using 10 or more Fe lines in the range 7300-8700~\AA, as listed in Table~\ref{linelist}. The metallicities of the sample stars  are listed in column~5 of Table~\ref{params}, along with their standard deviation.\\
\textit{Heavy elements}: 
\shree{The spectra of S stars are dominated by molecular bands and unblended atomic s-process lines are rare  \citep{smithlambertfeb1990}. This is why a spectral-synthesis approach is required, as opposed to relying solely on equivalent widths.}
As in \cite{pieter2015}, \cite{drisya} and S18, we only used the two \ion{Zr}{I} lines at 7819.37~\AA~and 7849.37~\AA~with transition probabilities from laboratory measurements \citep{Zrlines}. 
The \ion{Y}{I} and \ion{Y}{II} available lines lie in regions subject to blending by molecular lines of ZrO. The Y abundance of BD~+34$^\circ$1698 could not be derived, presumably because of its higher Zr abundance coupled with a low temperature ([Zr/Fe]=2.7, \teff=3400~K; see Tables \ref{params} and \ref{abundtable}). 
The Ba abundance was derived for all stars using the 7488.077~\AA~\ion{Ba}{I} line. 
We could not find any good Nd lines for BD~+34$^\circ$1698, HD~357941 and CSS~154, but for CSS~182 and CD~$-29^\circ$5912 at least 5 good lines were available. The other s-process element abundances were derived using the lines from Table~\ref{linelist} and are listed in Table~\ref{abundtable}.

All the sample stars show an overall mild enrichment of s-process elements. These results are also consistent with the moderate C/O ratios of these stars (0.5 -- 0.75), indicating that they are at the very beginning of the TP-AGB.

Finally, Table~\ref{abundtable} confirms the Tc~--~Nb anti-correlation encountered in S stars \citep{pieter2015}, since all the Tc-rich stars of our sample are devoid of any significant Nb enhancement, whereas they show genuine Zr overabundances. 
This is evidence
that the freshly produced $^{93}$Zr did not have time yet to decay to (mono-isotopic) Nb. Our classification (based on the Tc lines) of these objects as intrinsic TP-AGB stars is thus fully corroborated by the Nb diagnostic. 
\newline
\textit{Abundance uncertainties.}
The atmospheric parameters of S stars are unfortunately degenerated, in the sense that different combinations of $T_{\rm eff}$, \logg\ and C/O may lead to similar line strengths. \shree{ The effects of changing the stellar parameters may thus  compensate each other, so that the impact on abundances may be limited} 
(see the discussion in S18 and Sect.~\ref{Sect:degeneracy}). Therefore, quadratically adding the abundance uncertainties induced by each atmospheric-parameter variation within its uncertainty range would crudely overestimate the total error.
To evaluate the abundance uncertainties, we used instead the method of \cite{cayrel}.
This method involves finding a model (tagged `model H' in the list of Table~\ref{abundanderror}) that provides an almost equally good spectral fit as our best model (tagged `model A') in a representative spectral region (around \ion{Zr}{I} lines at 7819~\AA\ and 7849~\AA).
The difference between the abundances derived from models H and A, together with the line-to-line scatter and the error due to continuum placement (0.1~dex), were quadratically added to estimate the total uncertainty on each elemental abundance. 
When only one spectral line was available for a given element, a standard line-to-line scatter of 0.1~dex was adopted. The effect on the abundances of the variation of individual atmospheric parameters  can be found in Table~\ref{abundanderror}. 

\begin{figure*}
    \centering
    \includegraphics[scale=0.5,trim={2cm 1cm 1cm 1cm}]{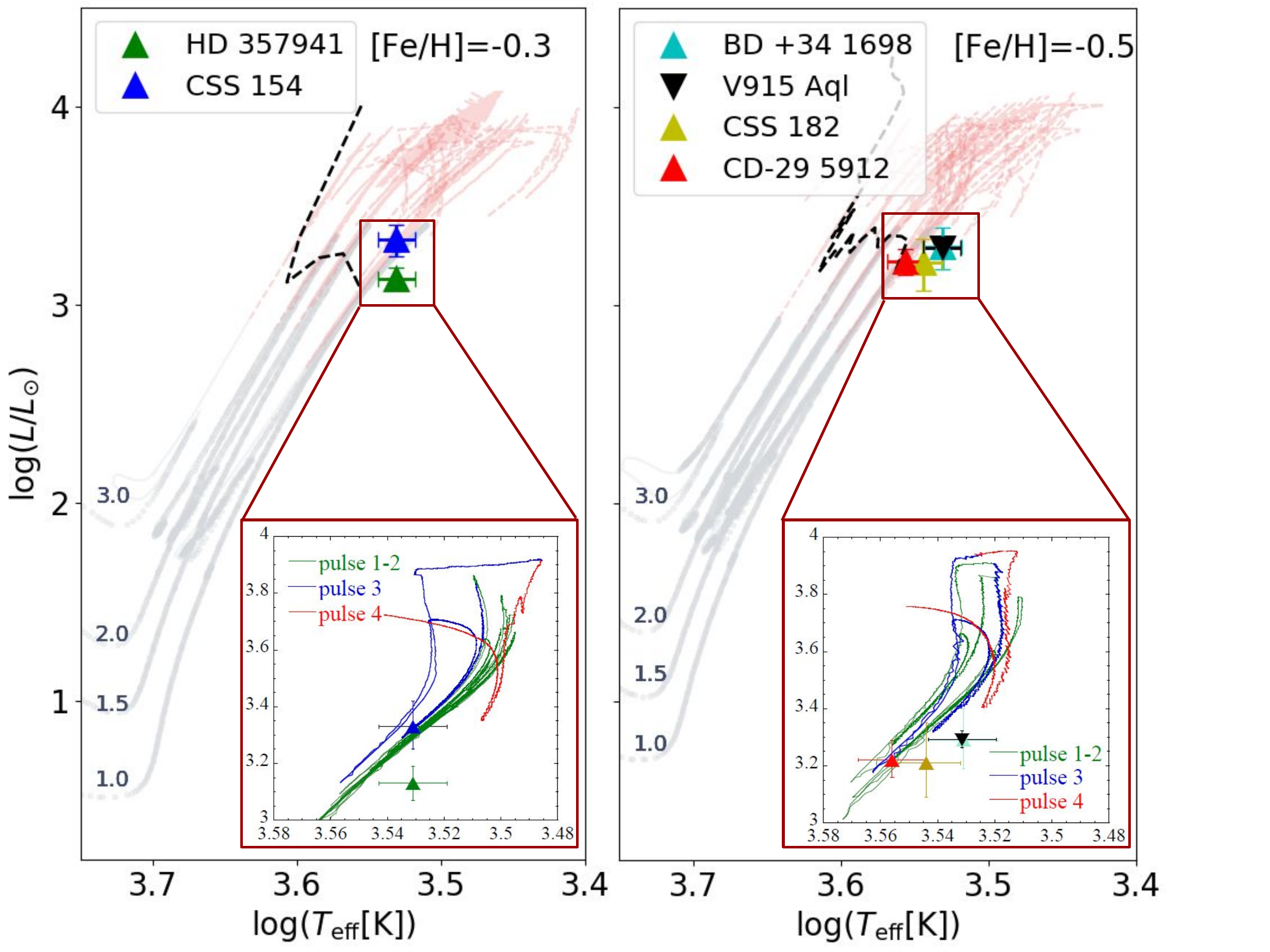}
    \caption{\label{HRD} Locations of the low-mass intrinsic S stars (triangles) in the HR diagram,
    compared with STAREVOL evolutionary tracks of the closest metallicity (as indicated on the top label) and of the labelled masses.
    In the large panels, the red giant branch and the core He-burning phases are depicted in grey, whereas the red dashed line corresponds to the AGB tracks. The black dashed line marks the predicted onset of TDU. In the insert, the different colors correspond to different \SG { pulse-interpulse cycles of the 1~\msun\ model star}, as labelled. 
    }
\end{figure*}


\section{Comparison with STAREVOL nucleosynthesis predictions}\label{abundancescomparison}
\shree{We computed low-mass AGB models in order to compare the measured s-process overabundances and C/O ratios with their predicted values. }
These AGB models have been generated with
the STAREVOL code \citep{siess2008} using an
extended s-process reaction network of 411 species and the same
input physics as described in \citet{goriely&siess}. 
The reference solar composition is taken from \citet{asplund2009} which corresponds to a metallicity $Z = 0.0134$. To describe
the mass-loss rate on the red giant branch (RGB), we use the \citet{Reimers1975} prescription
with $\eta_R=0.4$ \forreferee{(in Sect.~\ref{Sect:Reimers}, we evaluate the impact of this choice on the derived masses for our sample S stars)} from the main sequence up to the beginning
of the AGB and then switch to the \citet{Vassiliadis1993}
rate.  Dedicated models  \forreferee{with an initial mass of} 1~\msun\ have been computed for [Fe/H]~$=-0.3$ and $-0.5$.
In the present calculations, a diffusion equation is used to compute the partial mixing of protons in the C-rich layers at
the time of the TDU. Following the formalism of Eq. (9) of \citet{goriely&siess}, the diffusive mixing
parameters adopted in our simulations are $f_{\rm over}=0.14$, $D_{\rm min} = 10^7\,{\rm cm^2\, s^{-1}}$ and $p = 1/2$, where $f_{\rm over}$ controls the extent of the mixing, $D_{\rm min}$ the value of the diffusion coefficient at the innermost boundary of the diffusive region and $p$ is an additional free parameter describing the shape of the diffusion profile. It should be stressed that a careful study of the parameter space for $f_{\rm over}$, $D_{\rm min}$ and $p$ has been carried out and only the above-mentioned values have been found to give rise to early TDU episodes and s-process enrichments compatible with observations (see below). While TDU is definitely needed to ensure a proper surface enrichment for the star to become an S star, the diffusive mixing should be strong enough to produce a significant amount of s-elements. But, on the other hand, the diffusive mixing should not be too efficient, to avoid large s-overabundances in the relatively thin envelope of the 1~\msun\ models. More details will be given in a forthcoming paper.

\begin{figure*}
    \centering
    \includegraphics[scale=0.65,trim={3cm 1cm 3cm 1cm}]{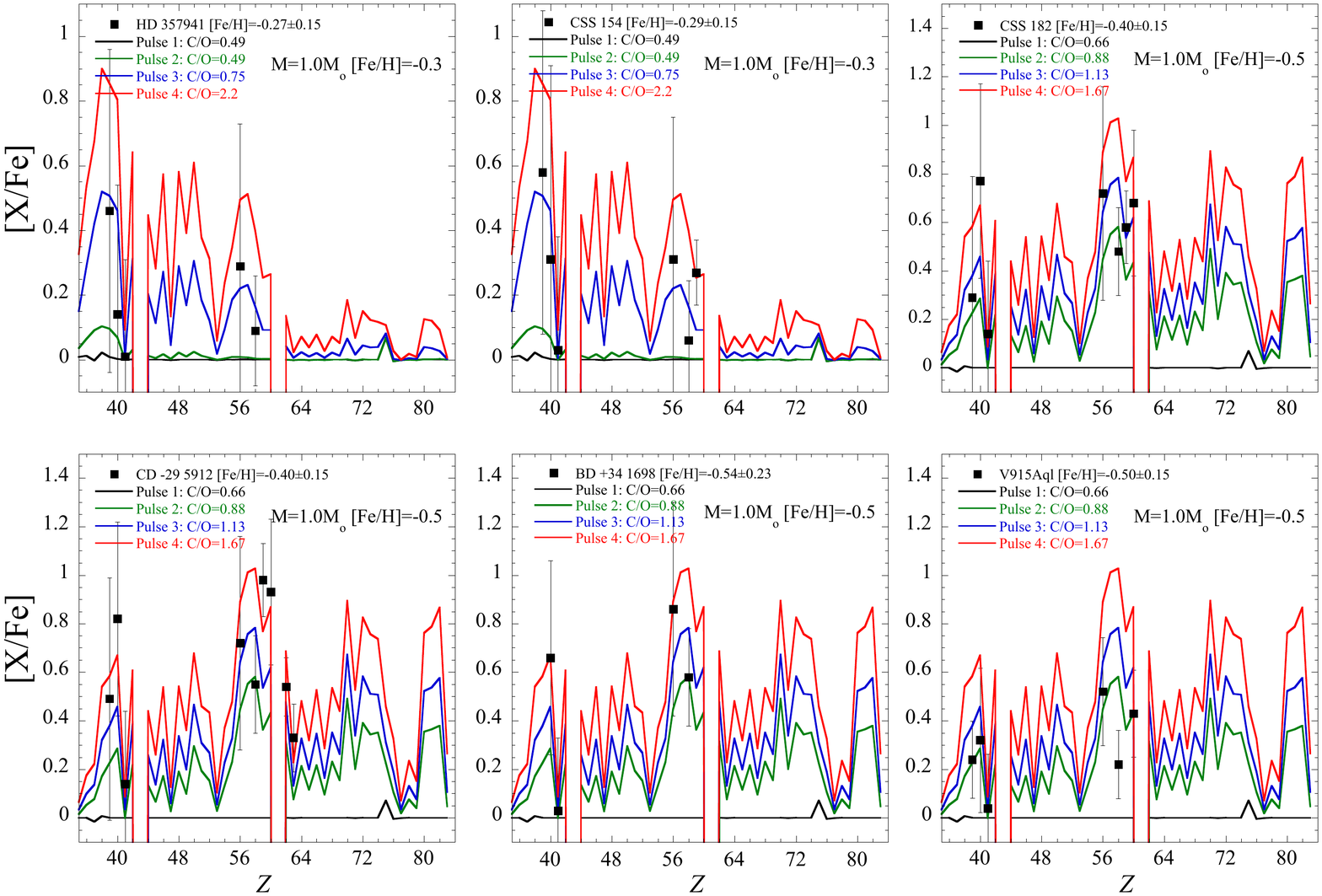}
    \caption{\label{nucleo}Comparison of the measured abundance profiles with the nucleosynthesis predictions.}
\end{figure*}

As shown in Fig.~\ref{HRD}, the location of our 6 stars in the HR diagram matches well the tracks corresponding to the  model star \forreferee{with initial mass 1~\msun\  (the estimated current mass is listed in the last column of Table~\ref{params})}, 
It can also be noted (see insert of Fig.~\ref{HRD}) that within the error bars on the effective temperatures and luminosities, observations are compatible with the theoretical tracks corresponding to the first 3 pulse-interpulse cycles 
of the 1~\msun\ model star.  
Note that an uncertainty of about 100~K on the model effective temperature should also be considered  \citep{cassisi2017}. The inclusion of an efficient diffusive mixing (with a relatively large value of $f_{\rm over}$) at the bottom of the stellar envelope triggers not only TDU at the end of the first fully developed thermal pulse, but also the mixing of protons into the C-rich layers, hence an s-process nucleosynthesis strong enough to account for the surface enrichment of our 6 stars. For those with  [Fe/H]~$\simeq -0.3$ (HD$\>$357941 and CSS$\>$154), the measured abundances displayed in Fig.~\ref{nucleo} are compatible with the occurrence of 3 thermal pulses, allowing the star to keep a relatively low C/O ratio (C/O = 0.75, compatible with observations; see Tables~\ref{params} and \ref{abundtable}). Among the lower-metallicity stars which are compared with [Fe/H]~$=-0.5$ models, CSS~182 and V915~Aql are compatible with a 2-pulse enrichment leading to C/O = 0.88. On the contrary, more TDU episodes seem to be required to explain the large surface abundances of s-elements measured in CD~$-29^\circ$5912 and BD~+34$^\circ$1698. \shree{However, these many TDU episodes then induce a C/O ratio above unity  which is incompatible with the measured C/O  ratios of 0.5 in these two objects.}

\section{Robustness of the derived stellar masses}
\label{Sect:degeneracy}

\forreferee{Here we evaluate the sensitivity of the derived stellar masses on two key factors: (i) the adopted atmospheric parameters (Sect.~\ref{Sect:degenerate_parameters}), and (ii) the mass-loss rate on the RGB (Sect.~\ref{Sect:Reimers}).
}

\subsection{Sensitivity to the adopted atmospheric-parameter set}
\label{Sect:degenerate_parameters}
The somewhat degenerate parameter space of S stars is well illustrated by the comparison of models A and H in Table~\ref{abundanderror}. For example, 
a change of \teff\ by +100~K or -100~K in model A can be compensated by an adjustment of the other stellar parameters (as in model H), in order to yield an equally acceptable fit of the global spectrum.
Since the location of the stars in the HR diagram is parameter-dependent,
so could in principle be the derived masses. 
The example of CD~$-29^\circ$5912, which is representative of the other stars, shows that this is not the case however
(Fig.~\ref{HRD_onlyCSS489}). 
Models A and H differ in \teff\ and C/O, and this difference induces a change in metallicity as derived from the Fe lines ([Fe/H]~$=-0.27$). However, when compared with evolutionary tracks of 
the corresponding 
 metallicity, the location in the HR diagram implied by models A or H in both cases falls along the track \forreferee{corresponding to the model with initial mass 1~\msun}.

\subsection{Sensitivity to the RGB mass-loss rate}
\label{Sect:Reimers}

\forreferee{The RGB mass-loss rate has a strong impact on the minimum initial mass for the occurrence of TDUs. Increasing the mass-loss rate by a factor of two along the RGB \citep[in agreement with the typical uncertainty on observed RGB mass-loss rates; see Table~8 of][]{schroder2007}, we find that a star with initial mass 1.2~\msun\ and $\eta_R = 0.8$ will reach the same TP-AGB location in the HR diagram as a 1~\msun\ model with the standard $\eta_R$ value of 0.4. This simple argument suggests to adopt an uncertainty of 0.2~\msun\ on the determination of the stellar mass on the basis of its location in the HR diagram, hence on the minimum initial mass for the occurrence of TDU.  }


\section{Conclusion}



The combination of dedicated MARCS model atmospheres with the Gaia DR2 parallaxes allows to derive  stellar parameters of S stars and to locate them in the HR diagram. Their \forreferee{(initial)} masses have been derived as well, comparing these locations with STAREVOL evolutionary tracks.
A subsample of 6 low-mass (1~--~1.2~\msun), intrinsic (Tc-rich, Nb-poor) S stars could be identified, and  points at the occurrence of TDU in low-mass AGB stars at relatively low luminosity, i.e. at the start of the TP-AGB.  
 Here, we present AGB models that can produce TDUs for \forreferee{initial} stellar masses as low as  1 \msun\ and metallicities [Fe/H] of $-0.3$ and $-0.5$, provided that a very efficient overshoot with $f_\mathrm{over}=0.14$ is applied at the base of the convective envelope. In the HR diagram, 
 our low-mass intrinsic S stars are nicely located just above the predicted onset of TDU events. 
 
 We obtain a reasonable agreement between the measured and predicted s-process  abundance profiles. However, some stars like CD~$-29^\circ$5912 and BD~+34$^\circ$1698 are puzzling since their level of s-process enrichment requires a number of TDU episodes that would bring the C/O ratio well above the measured value. Nevertheless, for the four remaining targets, the agreement between s-process abundances and C/O ratios is satisfactory.


\shree{In summary, our results on intrinsic low-mass S stars definitely prove that the TDU is active  in stars with masses as low as \forreferee{1~--~1.2~\msun} with [Fe/H] in the range $-0.3$ to $-0.5$. The new AGB STAREVOL models can now account for the measured s-process abundances of low-mass AGB stars. These new observations and models improve our understanding of the onset of TDU and its mass and metallicity dependence.}

\begin{acknowledgements} The authors thank the referee for very constructive comments.
      This research has been funded by the Belgian Science Policy Office under contract BR/143/A2/STARLAB. 
      S.V.E. thanks {\it Fondation ULB} for its support.
      Based on observations obtained with the HERMES spectrograph, which is supported by the Research Foundation - Flanders (FWO), Belgium, the Research Council of KU Leuven, Belgium, the \textit{Fonds National de la Recherche Scientifique} (F.R.S.-FNRS), Belgium, the Royal Observatory of Belgium, the \textit{Observatoire de Gen\`eve}, Switzerland and the \textit{Th\"{u}ringer Landessternwarte Tautenburg}, Germany. This work has made use of data from the European Space Agency (ESA)  mission
      Gaia \url{(https://www.cosmos.esa.int/gaia)},  processed  by the Gaia Data  Processing  and  Analysis  Consortium \url{(DPAC, https://www.cosmos.esa.int/web/gaia/dpac/consortium)}. Funding for the DPAC has
      been provided by national institutions, in particular the institutions participating in the Gaia
      Multilateral Agreement. This research has also made use of the SIMBAD database, operated at CDS, Strasbourg, France. LS \& SG are senior FNRS research associates. 
\end{acknowledgements}

\bibliographystyle{aa}
\bibliography{biblio}

\begin{thebibliography}{55}
\expandafter\ifx\csname natexlab\endcsname\relax\def\natexlab#1{#1}\fi

\bibitem[{{Alvarez} \& {Plez}(1998)}]{rodrigo}
{Alvarez}, R. \& {Plez}, B. 1998, \aap, 330, 1109

\bibitem[{{Arlandini} {et~al.}(1999){Arlandini}, {K{\"a}ppeler}, {Wisshak},
  {Gallino}, {Lugaro}, {Busso}, \& {Straniero}}]{arlandini1999}
{Arlandini}, C., {K{\"a}ppeler}, F., {Wisshak}, K., {et~al.} 1999, \apj, 525,
  886

\bibitem[{{Asplund} {et~al.}(2009){Asplund}, {Grevesse}, {Sauval}, \&
  {Scott}}]{asplund2009}
{Asplund}, M., {Grevesse}, N., {Sauval}, A.~J., \& {Scott}, P. 2009, ARA\&A,
  47, 481

\bibitem[{{Bi\'emont} {et~al.}(1981){Bi\'emont}, {Grevesse}, {Hannaford}, \&
  {Lowe}}]{Zrlines}
{Bi\'emont}, E., {Grevesse}, N., {Hannaford}, P., \& {Lowe}, R.~M. 1981, \apj,
  248, 867

\bibitem[{{Bisterzo} {et~al.}(2010){Bisterzo}, {Gallino}, {Straniero},
  {Cristallo}, \& {K{\"a}ppeler}}]{bisterzo}
{Bisterzo}, S., {Gallino}, R., {Straniero}, O., {Cristallo}, S., \&
  {K{\"a}ppeler}, F. 2010, \mnras, 404, 1529

\bibitem[{{Cassisi}(2017)}]{cassisi2017}
{Cassisi}, S. 2017, in European Physical Journal Web of Conferences, Vol. 160,
  04002

\bibitem[{{Cayrel} {et~al.}(2004){Cayrel}, {Depagne}, {Spite}, {Hill}, {Spite},
  {Fran{\c{c}}ois}, {Plez}, {Beers}, {Primas}, {Andersen}, {Barbuy},
  {Bonifacio}, {Molaro}, \& {Nordstr{\"o}m}}]{cayrel}
{Cayrel}, R., {Depagne}, E., {Spite}, M., {et~al.} 2004, \aap, 416, 1117

\bibitem[{{Corliss} \& {Bozman}(1962)}]{CB}
{Corliss}, C.~H. \& {Bozman}, W.~R. 1962, NBS Monograph, Vol.~53, {Experimental
  transition probabilities for spectral lines of seventy elements; derived from
  the NBS Tables of spectral-line intensities} (US Government Printing Office)

\bibitem[{{Cristallo} {et~al.}(2015){Cristallo}, {Straniero}, {Piersanti}, \&
  {Gobrecht}}]{cristallo}
{Cristallo}, S., {Straniero}, O., {Piersanti}, L., \& {Gobrecht}, D. 2015,
  ApJS, 219, 40

\bibitem[{{De Smedt} {et~al.}(2015){De Smedt}, {Van Winckel}, {Kamath}, \&
  {Wood}}]{2015kenneth}
{De Smedt}, K., {Van Winckel}, H., {Kamath}, D., \& {Wood}, P.~R. 2015, \aap,
  583, A56

\bibitem[{{Den Hartog} {et~al.}(2003){Den Hartog}, {Lawler}, {Sneden}, \&
  {Cowan}}]{HLSC}
{Den Hartog}, E.~A., {Lawler}, J.~E., {Sneden}, C., \& {Cowan}, J.~J. 2003,
  ApJS, 148, 543

\bibitem[{{Duquette} \& {Lawler}(1982)}]{DLa}
{Duquette}, D.~W. \& {Lawler}, J.~E. 1982, \pra, 26, 330

\bibitem[{{Fishlock} {et~al.}(2014){Fishlock}, {Karakas}, {Lugaro}, \&
  {Yong}}]{fishlock2014}
{Fishlock}, C.~K., {Karakas}, A.~I., {Lugaro}, M., \& {Yong}, D. 2014, \apj,
  797, 44

\bibitem[{{Gaia Collaboration} {et~al.}(2018){Gaia Collaboration}, {Brown},
  {Vallenari}, {Prusti}, {de Bruijne}, {Babusiaux}, {Bailer-Jones}, {Biermann},
  {Evans}, {Eyer}, \& et~al.}]{gdr2}
{Gaia Collaboration}, {Brown}, A.~G.~A., {Vallenari}, A., {et~al.} 2018, \aap,
  616, A1

\bibitem[{{Gontcharov}(2012)}]{gontcharov}
{Gontcharov}, G.~A. 2012, Astronomy Letters, 38, 87

\bibitem[{{Goriely} \& {Siess}(2018)}]{goriely&siess}
{Goriely}, S. \& {Siess}, L. 2018, \aap, 609, A29

\bibitem[{{Hannaford} \& {Lowe}(1985)}]{HL}
{Hannaford}, P. \& {Lowe}, R.~M. 1985, Journal of Physics B Atomic Molecular
  Physics, 18, 2365

\bibitem[{{Jorissen} {et~al.}(1993){Jorissen}, {Frayer}, {Johnson}, {Mayor}, \&
  {Smith}}]{jorissen1993}
{Jorissen}, A., {Frayer}, D.~T., {Johnson}, H.~R., {Mayor}, M., \& {Smith},
  V.~V. 1993, \aap, 271, 463

\bibitem[{{Jorissen} \& {Mayor}(1988)}]{jorissen1988}
{Jorissen}, A. \& {Mayor}, M. 1988, \aap, 198, 187

\bibitem[{{Karakas} \& {Lugaro}(2016)}]{karakas}
{Karakas}, A.~I. \& {Lugaro}, M. 2016, \apj, 825, 26

\bibitem[{{Karinkuzhi} {et~al.}(2018){Karinkuzhi}, {Van Eck}, {Jorissen},
  {Goriely}, {Siess}, {Merle}, {Escorza}, {Van der Swaelmen}, {Boffin},
  {Masseron}, {Shetye}, \& {Plez}}]{drisya}
{Karinkuzhi}, D., {Van Eck}, S., {Jorissen}, A., {et~al.} 2018, \aap, 618, A32

\bibitem[{{Kurucz}(2007)}]{K07}
{Kurucz}, R.~L. 2007, Robert L. Kurucz on-line database of observed and
  predicted atomic transitions

\bibitem[{{Lugaro} {et~al.}(2012){Lugaro}, {Karakas}, {Stancliffe}, \&
  {Rijs}}]{lugaro2012}
{Lugaro}, M., {Karakas}, A.~I., {Stancliffe}, R.~J., \& {Rijs}, C. 2012, \apj,
  747, 2

\bibitem[{{Martin} {et~al.}(1988){Martin}, {Fuhr}, \& {Wiese}}]{MFW}
{Martin}, G., {Fuhr}, J., \& {Wiese}, W. 1988, J. Phys. Chem. Ref. Data Suppl.,
  17

\bibitem[{{Mathews} {et~al.}(1986){Mathews}, {Takahashi}, {Ward}, \&
  {Howard}}]{Mathews1986}
{Mathews}, G.~J., {Takahashi}, K., {Ward}, R.~A., \& {Howard}, W.~M. 1986,
  \apj, 302, 410

\bibitem[{{Meggers} {et~al.}(1975){Meggers}, {Corliss}, \& {Scribner}}]{MC}
{Meggers}, W.~F., {Corliss}, C.~H., \& {Scribner}, B.~F. 1975, {Tables of
  spectral-line intensities. Part I, II - arranged by elements} (US Government
  Printing Office)

\bibitem[{{Merle} {et~al.}(2016){Merle}, {Jorissen}, {Van Eck}, {Masseron}, \&
  {Van Winckel}}]{tibaulth}
{Merle}, T., {Jorissen}, A., {Van Eck}, S., {Masseron}, T., \& {Van Winckel},
  H. 2016, \aap, 586, A151

\bibitem[{{Merrill}(1922)}]{Merrill}
{Merrill}, P.~W. 1922, \apj, 56, 457

\bibitem[{{Merrill}(1952)}]{merrill52}
{Merrill}, P.~W. 1952, \apj, 116, 21

\bibitem[{{Miles} \& {Wiese}(1969)}]{MW}
{Miles}, B.~M. \& {Wiese}, W.~L. 1969, Atomic Data, 1, 1

\bibitem[{{Nave} {et~al.}(1994){Nave}, {Johansson}, {Learner}, {Thorne}, \&
  {Brault}}]{NS}
{Nave}, G., {Johansson}, S., {Learner}, R.~C.~M., {Thorne}, A.~P., \& {Brault},
  J.~W. 1994, \apjs, 94, 221

\bibitem[{{Neyskens} {et~al.}(2015){Neyskens}, {Van Eck}, {Jorissen},
  {Goriely}, {Siess}, \& {Plez}}]{pieter2015}
{Neyskens}, P., {Van Eck}, S., {Jorissen}, A., {et~al.} 2015, \nat, 517, 174

\bibitem[{{Nilsson} {et~al.}(1991){Nilsson}, {Johansson}, \& {Kurucz}}]{Nil}
{Nilsson}, A.~E., {Johansson}, S., \& {Kurucz}, R.~L. 1991, \physscr, 44, 226

\bibitem[{{O'Brian} {et~al.}(1991){O'Brian}, {Wickliffe}, {Lawler}, {Whaling},
  \& {Brault}}]{BWL}
{O'Brian}, T.~R., {Wickliffe}, M.~E., {Lawler}, J.~E., {Whaling}, W., \&
  {Brault}, J.~W. 1991, Journal of the Optical Society of America B Optical
  Physics, 8, 1185

\bibitem[{{Palmeri} {et~al.}(2000){Palmeri}, {Quinet}, {Wyart}, \&
  {Bi{\'e}mont}}]{PQWB}
{Palmeri}, P., {Quinet}, P., {Wyart}, J., \& {Bi{\'e}mont}, E. 2000, Physica
  Scripta, 61, 323

\bibitem[{{Plez}(2012)}]{turbospectrum}
{Plez}, B. 2012, {Turbospectrum: Code for spectral synthesis}, Astrophysics
  Source Code Library

\bibitem[{{Raskin} {et~al.}(2011){Raskin}, {van Winckel}, {Hensberge},
  {Jorissen}, {Lehmann}, {Waelkens}, {Avila}, {de Cuyper}, {Degroote},
  {Dubosson}, {Dumortier}, {Fr{\'e}mat}, {Laux}, {Michaud}, {Morren}, {Perez
  Padilla}, {Pessemier}, {Prins}, {Smolders}, {Van Eck}, \& {Winkler}}]{raskin}
{Raskin}, G., {van Winckel}, H., {Hensberge}, H., {et~al.} 2011, \aap, 526, A69

\bibitem[{{Reimers}(1975)}]{Reimers1975}
{Reimers}, D. 1975, M\'emoires de la Soci\'et\'e Royale des Sciences de
  Li\`ege, 8, 369

\bibitem[{{Schr{\"o}der} \& {Cuntz}(2007)}]{schroder2007}
{Schr{\"o}der}, K.-P. \& {Cuntz}, M. 2007, \aap, 465, 593

\bibitem[{{Shetye} {et~al.}(2018){Shetye}, {Van Eck}, {Jorissen}, {Van
  Winckel}, {Siess}, {Goriely}, {Escorza}, {Karinkuzhi}, \& {Plez}}]{shreeya1}
{Shetye}, S., {Van Eck}, S., {Jorissen}, A., {et~al.} 2018, \aap, 620, A148

\bibitem[{{Siess} \& {Arnould}(2008)}]{siess2008}
{Siess}, L. \& {Arnould}, M. 2008, \aap, 489, 395

\bibitem[{{Smith} \& {Lambert}(1988)}]{smith1988}
{Smith}, V.~V. \& {Lambert}, D.~L. 1988, \apj, 333, 219

\bibitem[{{Smith} \& {Lambert}(1990)}]{smithlambertfeb1990}
{Smith}, V.~V. \& {Lambert}, D.~L. 1990, \apjs, 72, 387

\bibitem[{{Stancliffe} {et~al.}(2005){Stancliffe}, {Izzard}, \&
  {Tout}}]{stancliffe}
{Stancliffe}, R.~J., {Izzard}, R.~G., \& {Tout}, C.~A. 2005, \mnras, 356, L1

\bibitem[{{Stephenson}(1984)}]{cgss}
{Stephenson}, C.~B. 1984, Publications of the Warner \& Swasey Observatory, 3,
  1

\bibitem[{{Straniero} {et~al.}(2003){Straniero}, {Dom{\'\i}nguez}, {Cristallo},
  \& {Gallino}}]{straniero}
{Straniero}, O., {Dom{\'\i}nguez}, I., {Cristallo}, S., \& {Gallino}, R. 2003,
  PASA, 20, 389

\bibitem[{{van Aarle} {et~al.}(2013){van Aarle}, {Van Winckel}, {De Smedt},
  {Kamath}, \& {Wood}}]{vanaarle}
{van Aarle}, E., {Van Winckel}, H., {De Smedt}, K., {Kamath}, D., \& {Wood},
  P.~R. 2013, \aap, 554, A106

\bibitem[{{Van Eck} \& {Jorissen}(1999)}]{VanEck-1999}
{Van Eck}, S. \& {Jorissen}, A. 1999, \aap, 345, 127

\bibitem[{{Van Eck} \& {Jorissen}(2000)}]{VanEck-2000b}
{Van Eck}, S. \& {Jorissen}, A. 2000, \aap, 360, 196

\bibitem[{{Van Eck} {et~al.}(2000){Van Eck}, {Jorissen}, {Udry}, {Mayor},
  {Burki}, {Burnet}, \& {Catchpole}}]{VanEck-2000a}
{Van Eck}, S., {Jorissen}, A., {Udry}, S., {et~al.} 2000, \aaps, 145, 51

\bibitem[{{Van Eck} {et~al.}(2017){Van Eck}, {Neyskens}, {Jorissen}, {Plez},
  {Edvardsson}, {Eriksson}, {Gustafsson}, {J{\o}rgensen}, \&
  {Nordlund}}]{sophieMarcs}
{Van Eck}, S., {Neyskens}, P., {Jorissen}, A., {et~al.} 2017, \aap, 601, A10

\bibitem[{{Vassiliadis} \& {Wood}(1993)}]{Vassiliadis1993}
{Vassiliadis}, E. \& {Wood}, P.~R. 1993, \apj, 413, 641

\bibitem[{{Wang} \& {Chen}(2002)}]{wang}
{Wang}, X.~H. \& {Chen}, P.~S. 2002, \aap, 387, 129

\bibitem[{{Weiss} \& {Ferguson}(2009)}]{weiss2009}
{Weiss}, A. \& {Ferguson}, J.~W. 2009, \aap, 508, 1343

\bibitem[{{Wenger} {et~al.}(2000){Wenger}, {Ochsenbein}, {Egret}, {Dubois},
  {Bonnarel}, {Borde}, {Genova}, {Jasniewicz}, {Lalo{\"e}}, {Lesteven}, \&
  {Monier}}]{simbad}
{Wenger}, M., {Ochsenbein}, F., {Egret}, D., {et~al.} 2000, \aaps, 143, 9

\end{thebibliography}

\begin{appendix}
\addappheadtotoc

\section{[s/Fe] in the MARCS grid}
\forreferee{
As a complement to \cite{sophieMarcs}, we give here details about the exact meaning of the [s/Fe] parameter in the MARCS grid of S stars, which provides models with [s/Fe] = 0, +1, and +2 dex. This parameter adjusts the abundances $\log \epsilon_X$ for elements from Ga to Bi in the following way: \\
$X_s = \log_{10}\left[ (n_{\rm X}/n_{\rm H})_\odot \times 10^\mathrm{[Fe/H]} \times 10^{[\mathrm{s}/\mathrm{Fe}]}  \times f_{X,s} \right] \;,$\\
$X_r = \log_{10}\left[(n_{\rm X}/n_{\rm H})_\odot \times 10^\mathrm{[Fe/H]} \times 10^{[\mathrm{r}/\mathrm{Fe}]}  \times (1 - f_{X,s})\right]  \;,$\\
$\log \epsilon_X = \log_{10}(10^{X_s} + 10^{X_r}$). \\
The fractional s-process abundance of element $X$ (denoted $f_{X,s}$) is taken from \cite{arlandini1999}. The fractional r-process contribution of that element is then simply given by $f_{X,r} = 1-f_{X,s}$. The parameter [r/Fe] is included in the above formulae for the sake of completeness, but it is set to 0 in the S-star grid. 
}

\clearpage
\newpage

\section{Basic data, atmospheric parameters and abundances of our sample stars}
\setcounter{table}{0}
\renewcommand{\thetable}{B\arabic{table}}

\setcounter{figure}{0}
\renewcommand{\thefigure}{B\arabic{figure}}

\begin{table*}
\caption{\label{basic data} Basic data of our sample stars. Column~1 lists the entry number in  the General Catalog of S Stars \citep{cgss}, and column~2 lists various other  identifiers. Column~3 gives the observation date and column~4 the S/N ratio in the $V$ band (around 500 nm). Columns~5, 6, 7, and 8 list the spectral type, the $V$ magnitude, the $V-K$ color index, and the 2MASS $K$ magnitude, respectively, retrieved from the SIMBAD database \citep{simbad}. Column~9 displays the Gaia DR2 parallax and its error. Column~10 lists the adopted reddening \citep{gontcharov}.}
\begin{center}

\begin{tabular}{l l l c l c c c c c c}
\hline
 CSS & Name & Observation date & S/N & Sp. type & $V$ & $V-K$ & $K$ & \multicolumn{1}{c}{$\varpi\pm\sigma_{\varpi}$} & $E_{B-V}$&  \\
  &  &  & & &  &  &  & (mas) & &  \\
\hline
1190 & HD 357941& 17 June 2018 & 90 & M4S  & 9.41 & 6.636 & 2.774 & 1.71 $\pm$ 0.11 & 0.057 &  \\
154 & IRAS 05387+0137 &  2 February 2017 & 40 & S5*3 & 10.79 &  6.997 & 3.793 & 0.85 $\pm$ 0.08 & 0.138 & \\
182 & IRAS 06000+1023 & 3 February 2017 & 60 & S4 & 10.38 & 6.119 & 4.261 & 0.81 $\pm$ 0.12 & 0.066 & \\
489 & CD $-29 ^\circ$5912& 17 February 2018 & 50 & S4, 4 & 10.79 & 5.751 & 5.039 & 0.59 $\pm$ 0.04 & 0.162 & \\
1099 & V915~Aql & 27 May 2016 & 50 & S5+/2 & 8.4 & 6.3 & 2.1 & 1.97 $\pm$ 0.06 & 0.17 & \\
413 & BD $+34 ^\circ$1698 & 23 April 2016 & 40 & M4S  & 10.67 & 6.965 & 3.705 & 0.93 $\pm$ 0.11 & 0.102 &  \\
\hline
\end{tabular}
    
\end{center}
\end{table*}

\begin{table*}
\setlength\tabcolsep{2pt}
\caption{Atmospheric parameters of the sample stars. The values in parentheses list the range spanned by the \logg~iterations (Sect.~\ref{sectionparams}), except for [Fe/H], where they list the uncertainties on [Fe/H] as derived from the abundance analysis (Sect.~\ref{abundsection}). \forreferee{ The last column provides the current mass of the S star, provided by the evolutionary track going through the current \teff\  and  luminosity of the star, for the corresponding metallicity. In all cases, the initial mass was 1~\msun.}}
\label{params}
    \centering
    \begin{tabular}{lccccccc}
    \hline
    Name & \teff & $L$ & \logg  & [Fe/H]  & C/O & [s/Fe] & \forreferee{M$_{\rm cur}$}\\
     & (K) & (L$_\odot$) &  &  & & (dex) & \forreferee{(\msun)} \\ 
     \hline
      HD 357941 & 3400 & 1357 &1 & -0.27   & 0.5 & 0 & \forreferee{0.7}\\
       &(100; 100) &(1193; 1558) &(0; 1) & ($\pm$ 0.23)  &(0.5; 0.75) & (0; 1)& \\
      CSS 154 & 3400 & 2128 & 1 & -0.29   & 0.5 & 0 & \forreferee{0.9}\\
       &(100; 100) & (1774; 2599) &(1; 3) & ($\pm$0.20) &(0.5; 0.89) & (0; 1)&\\
      CSS 182 & 3500 &1635& 1 & -0.4   & 0.5 & 1 & \forreferee{0.9}\\
       &(100; 100)&(1236; 2266) &(1; 3) & ($\pm$0.21)  &(0.5; 0.89) & (1; 1)&\\
      CD $-29^\circ$5912 & 3600 & 1667 & 1 & -0.4   & 0.5 & 1 & \forreferee{0.8}\\
       &(100; 100) & (1446; 1943)& (1; 3) & ($\pm$0.22)  &(0.5; 0.89) & (1; 1)&\\
       V915 Aql       & 3400  & 1958  & 0  & -0.5  & 0.75 & 0  & \forreferee{0.7} \\
 & (3400; 3400) & (1832; 2098) & (0; 1) & ($\pm$ 0.15) & (0.65; 0.75) & (0; 1)& \\
    BD +34$^\circ$1698 & 3400 & 1967 &1 & -0.54   & 0.5 & 1 & \forreferee{0.7}\\
      &(200; 200)& (1564; 2548) & (1; 3) &($\pm$0.27)  &(0.5; 0.89)&(1; 1) &\\
    \hline
    
    \end{tabular}
    
\end{table*}

\begin{table*}[]
\caption{\label{abundtable}Elemental abundances for the sample stars, along with the standard deviation due to line-to-line scatter. Solar abundances (third column) are from \cite{asplund2009}. The  column labelled $N$ lists the number of lines used to derive the abundance. The $\sigma_{\rm [X/Fe]}$ column lists the total uncertainty on the abundances calculated using the method described in Sect.~\ref{abundsection}. The abundances and total error budget for V915~Aql have been retrieved from S18.}
\setlength\tabcolsep{1.5pt}
\begin{tabular}{lcc|ccccc|ccccc|ccccc}
\hline
 & & &\multicolumn{5}{c}{BD +34$^\circ$1698}&\multicolumn{5}{c}{HD 357941}& \multicolumn{5}{c}{CSS 154} \\
\hline
      & $Z$  & $\log {\epsilon^a}_\odot$ & $\log \epsilon$  & $N$       & {[}X/H{]} & {[}X/Fe{]}& $\sigma_{\rm [X/Fe]}$ & $\log \epsilon$  & $N$        & {[}X/H{]} & {[}X/Fe{]}& $\sigma_{\rm[X/Fe]}$ & $\log \epsilon$  & $N$       & {[}X/H{]} & {[}X/Fe{]} & $\sigma_{\rm[X/Fe]}$\\
\hline
C     & 6  & 8.43  & 8.059      &  -&  -0.371 & 0.169 &- & 8.059      &-&   -0.371        & -0.101 &-  &  8.359     & -&    -0.071  & 0.219 &- \\
N     & 7  & 7.83  & 9.6      &   -      &  1.77  & 2.31   & -       & 9.1      &  -        &    1.27   &  1.54&-  & 9.6      &      -   & 1.77      &  2.06 & -\\
O     & 8  & 8.69  & 8.36 &    - &  -0.33       &   0.21   & -      & 8.36      & -         & -0.33          & -0.06 &-    & 8.66      & -        & -0.03 &  0.26 & - \\
Fe    & 26 & 7.5   & 6.96 $\pm0.27$  & 13      & -0.54     &-& 0.3 & 7.23 $\pm0.23$  & 13       & -0.27     & - &0.2 & 7.21$\pm0.20$  & 10      & -0.29     &      - &   0.2  \\
Y I   & 39 & 2.21  &-&-&-&-&- & 2.3$\pm$ 0.00    & 1        & 0.09      & 0.36  & 0.5    & 2.5$\pm$ 0.00    & 1 & 0.29& 0.58   & 0.5   \\
Y II  & 39 & 2.21  &-&-&-&-&-& 2.5$\pm$ 0.00   & 1        & 0.29      & 0.56 &  0.5    &  -& - &  -&-&- \\
Zr I  & 40 & 2.58  & 2.7$\pm$ 0.07   & 2       & 0.12      & 0.66 & 0.4 & 2.45 $\pm$ 0.21 & 2        & -0.13     & 0.14 &0.4  & 2.6$\pm$ 0.42   & 2       & 0.02      & 0.31  & 0.6    \\
Nb I  & 41 & 1.46  & 0.95  $\pm$ 0.05  & 3       & -0.51     & 0.03 &  0.3  & 1.2$\pm$  0.00    & 2        & -0.26     & 0.01& 0.3  & 1.2 $\pm$ 0.00     & 1       & -0.26     & 0.03 & 0.3  \\
Ba I  & 56 & 2.18  & 2.5$\pm$ 0.00    & 1       & 0.29      & 0.86 &0.4 & 2.2$\pm$ 0.00   & 1        & 0.02      & 0.29  &0.4 & 2.2  $\pm$ 0.00      & 1       & 0.02      & 0.31 &  0.4 \\
Ce II & 58 & 1.58   & 1.62$\pm$ 0.15  & 4       & 0.04     &   0.58  &  0.2    & 1.4  $\pm$ 0.00  & 2        & -0.18      & 0.09 & 0.2 & 1.35 $\pm$ 0.07   & 2       & -0.23     & 0.06 &  0.2  \\
Pr II & 59 & 0.72  &-& -& -&-& -&-&-& -&-& -& 0.7 $\pm$ 0.00   & 1 & -0.02     & 0.27 &  0.1  \\
\hline\\
\end{tabular}

\begin{center}
\begin{tabular}{l c c | c c c c c| c c c c c| c c c c c}
\hline
 & & &\multicolumn{5}{c}{CSS 182}&\multicolumn{5}{c}{CD $-29^\circ5912$}&\multicolumn{5}{c}{V915~Aql}\\
\hline
      & $Z$  & $\log {\epsilon^a}_\odot$ & $\log \epsilon$  & $N$       & {[}X/H{]} & {[}X/Fe{]}& $\sigma_{\rm [X/Fe]}$& $\log \epsilon$  & $N$       & {[}X/H{]} & {[}X/Fe{]}& $\sigma_{\rm [X/Fe]}$ & $\log \epsilon$  & $N$       & {[}X/H{]} & {[}X/Fe{]}&$\sigma_{\rm [X/Fe]}$   \\
\hline
C     & 6  & 8.43  & 8.059       &  - &    -0.371       & 0.029&-  &  8.059     & - & -0.371  &  0.029&-  & 8.24 &- & -0.19 & 0.31 &-\\
N     & 7  & 7.83  & 9.3     &  -&    1.47       &    1.87  &-      &  8.9     &  -&  1.07         & 1.47 &-          & 7.60  &- & -0.2 & 0.3&-\\
O     & 8  & 8.69  & 8.36      &-&   -0.33        &     0.07 &-      & 8.36      &-& -0.33           & 0.07 &-      & 8.36  & -&-0.33 & 0.17 &-    \\
Fe    & 26 & 7.5   & 7.1 $\pm0.21$   & 13      & -0.4      &-&0.2 & 7.1 $\pm0.22$  & 18      & -0.4      & - &0.2 & 7.0 $\pm$0.16& 10 & -0.50 &-  &                        0.2 \\
Y I   & 39 & 2.21  & 2.06  $\pm$0.2  & 5       & -0.15     & 0.25 &0.6 & 2.3 $\pm$ 0.18  & 4       & 0.09      & 0.49 &0.5 & 1.9 $\pm$0.00 &1& -0.31 & 0.19 & 0.1 \\
Y II  & 39 & 2.21  & 2.5    $\pm$ 0.00 & 1       & 0.29      & 0.69  &0.5 & -& -&  -&  -&-& 2.0 $\pm$ 0.00 &1&-0.21&0.29 & 0.2 \\
Zr I  & 40 & 2.58  & 2.95 $\pm$ 0.07 & 2       & 0.37      & 0.77 &0.4  & 3.0 $\pm$ 0.07     & 2       & 0.42      & 0.82 &0.4 & 2.4 $\pm$ 0.28 & 2& -0.18 & 0.32 & 0.2 \\
Nb I  & 41 & 1.46  & 1.2  $\pm$0.00 & 2       & -0.26     & 0.14 & 0.3  & 1.2 $\pm$ 0.00   & 2       & -0.26     & 0.14 &0.3 & 1.00$\pm$0.00 & 1&-0.46 & 0.04& 0.1\\
Ba I  & 56 & 2.18  & 2.5  $\pm$0.00 & 1       & 0.32      & 0.72 & 0.4  & 2.5 $\pm$ 0.00   & 1       & 0.32      & 0.72 &0.4  & 2.2$\pm$0.00 &1& 0.02 & 0.52&0.2\\
Ce II & 58 & 1.58  & 1.66 $\pm$0.08 & 5       & 0.08     & 0.48 &0.2  & 1.73 $\pm$ 0.05 & 3       & 0.15      & 0.55 &0.2 & 1.30 $\pm$0.00&3 & -0.28 & 0.22 &0.1  \\
Pr II & 59 & 0.72  & 0.9  $\pm$0.00 & 1       & 0.18      & 0.58& 0.1  & 1.3  $\pm$ 0.00 & 1       & 0.58      & 0.98&0.1  &-&-&-&-&- \\
Nd II & 60 & 1.42  & 1.7  $\pm$0.32 & 5       & 0.28      & 0.68&  0.3 & 1.95   $\pm$ 0.25& 6       & 0.53      & 0.93 &0.3& 1.35 $\pm$ 0.14 &2&-0.07 & 0.43&0.2\\
Sm II & 62 & 0.96  &- & -&  -& -&-& 1.1   $\pm$ 0.00  & 1       & 0.14      & 0.54 & 0.1&-&-&-&-&-\\
Eu II & 63 & 0.52  &- &-& -&-&-&0.45 $\pm$ 0.07  & 2       & -0.07     & 0.33&0.1  &- & -& -& -&-\\
\hline
\end{tabular}
\end{center}
\end{table*}

\begin{table}[]
\begin{centering}
\setlength\tabcolsep{4pt}
\caption{\label{abundanderror} Sensitivity of the elemental abundances of CD $-29^\circ$5912 on variations of the atmospheric parameters. A dash in the $\Delta$ columns indicates that because of a degraded agreement between the observed and the synthetic spectrum, the (unique) line usually providing the abundance for the considered element had to be rejected.}
\begin{tabular}{cccrlcc}
\hline
model & $T_{\rm eff}$ & $\log g$ & {[}Fe/H{]} & C/O   & {[}s/Fe{]} & $\chi_{t}$ \\
      & (K)  & (dex) & (dex)      &       & (dex)      & (km~s $^{-1}$) \\
\hline
A     & 3600 & 1     & -0.5       & 0.5   & 1          & 2      \\
B     & 3700 & 1     & -0.5       & 0.5   & 1          & 2      \\
C     & 3500 & 1     & -0.5       & 0.5   & 1          & 2      \\
D     & 3600 & 0     & -0.5       & 0.5   & 1          & 2      \\
E     & 3600 & 1     & -0.5       & 0.5   & 1          & 1.5    \\
F     & 3600 & 1     & -0.5       & 0.75  & 1          & 2      \\
G     & 3600 & 1     & 0.0          & 0.5   & 1          & 2      \\
H     & 3500 & 1     & -0.5       & 0.9 & 1          & 2     \\
\hline\\
\end{tabular}

\begin{center}
\begin{tabular}{lrrrrrrr}
\hline
Element     & $\Delta_{B-A}$           & $\Delta_{C-A}$  & $\Delta_{D-A}$          & $\Delta_{E-A}$          & $\Delta_{F-A}$            & $\Delta_{G-A}$            &$\Delta_{H-A}$           \\
\hline
{[}N/Fe{]}  & -0.54 & 0.27  & -0.71 & 0.05  & -0.75 & -0.14 & -1.28         \\
{[}Fe/H{]}  & -0.02 & 0.20  & -0.20 & 0.21  & 0.04  & 0.28  & 0.13          \\
{[}Y/Fe{]}  & 0.16  & -0.10 & -0.07 & 0.02  & -0.19 & -0.02 & -0.56         \\
{[}Zr/Fe{]} & -0.13 & -0.20 & -0.05 & -0.11 & -0.14 & 0.22  & -0.43         \\
{[}Nb/Fe{]} & 0.32  & -0.20 & -     & -0.11 & -0.04 & 0.02  & -0.33         \\
{[}Ba/Fe{]} & 0.02  & -0.50 & -0.10 & -0.51 & -0.34 & -0.28 & -0.43         \\
{[}Ce/Fe{]} & 0.02  & -0.20 & -0.23 & 0.03  & -0.07 & 0.18  & -0.19         \\
{[}Pr/Fe{]} & 0.22  & -     & -0.10 & -     & -0.04 & -     & -0.03         \\
{[}Nd/Fe{]} & 0.02  & -0.20 & -0.03 & 0.16  & -0.04 & 0.11  & -0.19 \\
{[}Sm/Fe{]} & 0.02  & -0.20 & -0.10 & -0.11 & 0.06  & -0.08 & -0.03 \\
{[}Eu/Fe{]} & 0.17  & -     & -0.10 & -0.16 & 0.01  & -0.23 & -0.08 \\
\hline
\end{tabular}
\end{center}
\end{centering}
\end{table}

\clearpage
\newpage
\section{Example spectra}
\setcounter{table}{0}
\renewcommand{\thetable}{C\arabic{table}}

\setcounter{figure}{0}
\renewcommand{\thefigure}{C\arabic{figure}}

\begin{figure*}      
\begin{centering}
    \mbox{\includegraphics[scale=0.42,trim={0cm 0cm 6cm 0cm}]{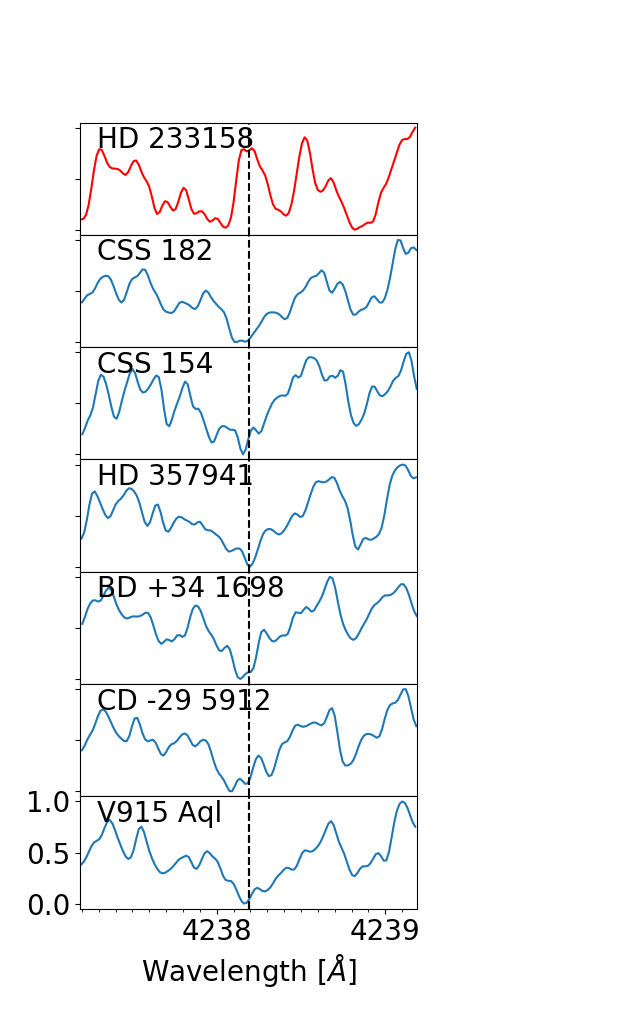}}   
    \hspace{0px}
    \mbox{\includegraphics[scale=0.42,trim={0cm 0cm 6cm 0cm}]{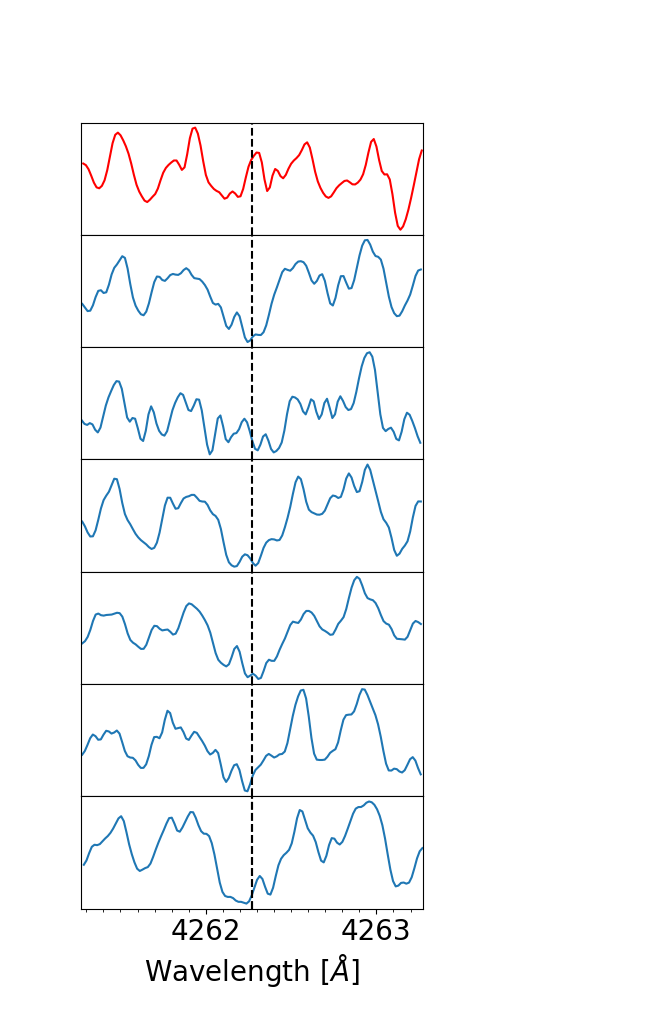}}
    \hspace{0px}
    \mbox{\includegraphics[scale=0.42,trim={0cm 0cm 6cm 0cm}]{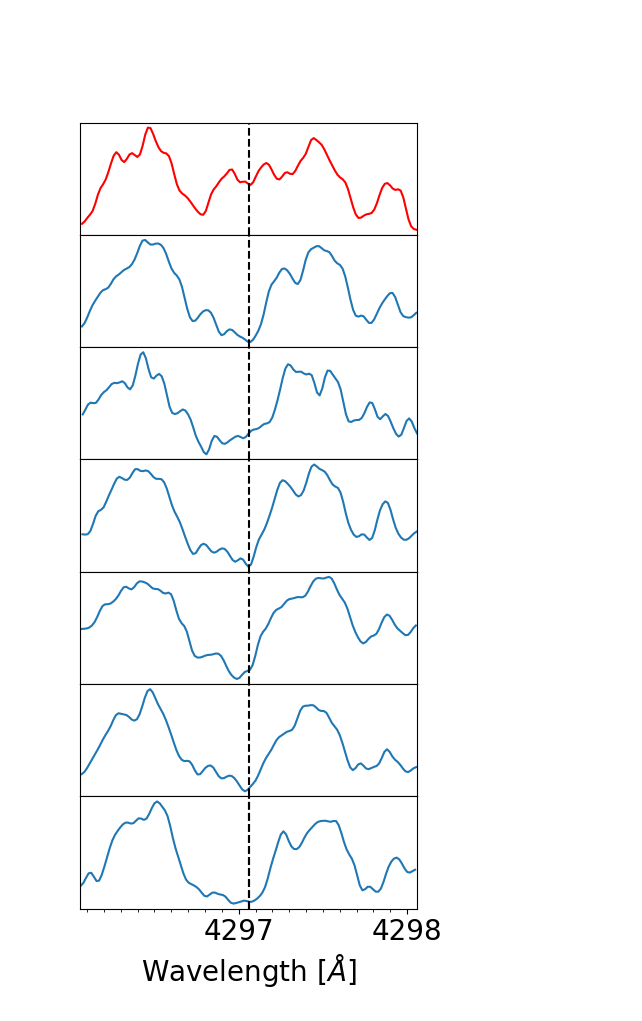}}
    \caption{\label{Tc} The spectral region around the three
    (4238.19, 4262.27 and 4297.06~\AA) violet \ion{Tc}{I} lines in the sample S stars as well as in a Tc-poor S star from S18  (HD 233158, red curve in the top panels), for the sake of comparison. The spectra have been arbitrarily normalized and binned by a factor of 1.5 to increase the S/N ratio.  }
\end{centering}
\end{figure*}

\begin{figure}      
\begin{centering}
    \includegraphics[scale=0.41,trim={1cm 1cm 2cm 1cm}]{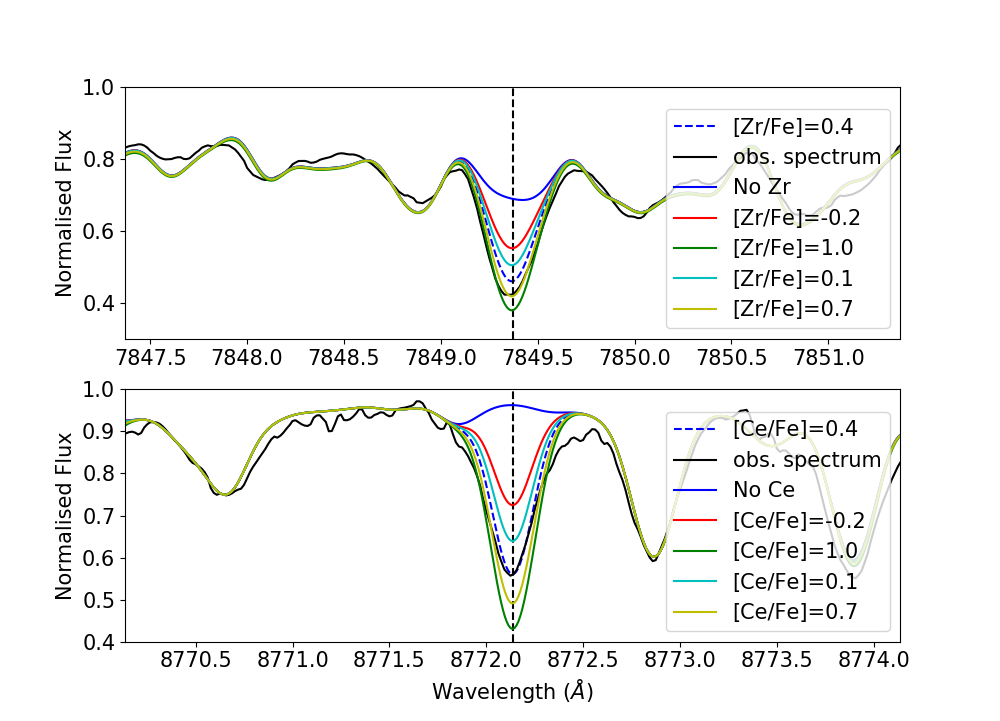}
    \caption{\label{s-processfits} Illustration of the quality of the match between observed and synthetic spectra obtained for the S star CSS~182 around the \ion{Zr}{I} line at 7849.37 \AA~(upper panel) and the \ion{Ce}{II} line at 8772.16 \AA~(lower panel).}
  
    \label{materialflowChart}
\end{centering}
\end{figure}

\clearpage
\newpage

\section{HR diagram}
\setcounter{table}{0}
\renewcommand{\thetable}{D\arabic{table}}

\setcounter{figure}{0}
\renewcommand{\thefigure}{D\arabic{figure}}

\begin{figure}[h!]
    \centering
    \includegraphics[scale=0.5]{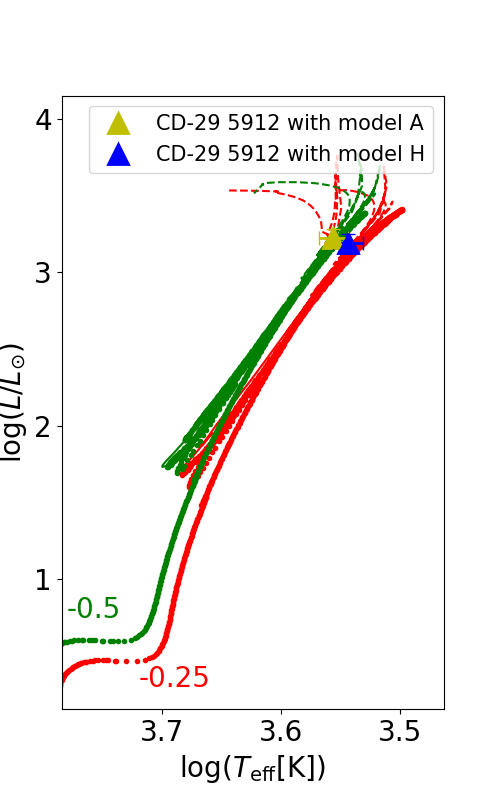}
    \caption{Location of CD $-29^\circ$5912 in the HR diagram with parameters of model A (\teff\ = 3600~K, $\log g =1$, [Fe/H] = $-0.4$, C/O = 0.5, [s/Fe] = 1),  and model H (\teff = 3500~K, $\log g =1$, [Fe/H] = $-0.27$, C/O = 0.9, [s/Fe] = 1) compared with the 1 \msun\ evolutionary tracks of the closest metallicities (labelled with the same color as the track). }
    \label{HRD_onlyCSS489}
\end{figure}

\clearpage
\newpage
\section{Linelist}
\setcounter{table}{0}
\renewcommand{\thetable}{E\arabic{table}}

\setcounter{figure}{0}
\renewcommand{\thefigure}{E\arabic{figure}}

\begin{table*}
\centering
\caption{Atomic line list. The last column identifies the stars where the corresponding line was used for abundance determination (A: HD~357941; B: CSS~154; C: CSS~182; D: CD $-29^\circ$5912; E: BD +34$^\circ$1698).}
\label{linelist}
\begin{tabular}{c c c c c c}

\hline
Species& $\lambda$ [\AA] & $\chi$ [eV] & $\log gf$ & Reference & Star\\
\hline
Fe I & 7389.398 & 4.301 & -0.460 & \cite{K07} & ACDE\\
 & 7418.667 & 4.143 & -1.376 & \cite{BWL} & ABCDE\\
 & 7443.022 & 4.186 & -1.820 & \cite{MFW} & ABCD\\
 & 7461.263 & 5.507 & -3.059 & \cite{K07} & ABCDE\\
 & 7498.530 & 4.143 & -2.250 & \cite{MFW} & ABCDE\\
 & 7540.430 & 2.727 & -3.850 & \cite{MFW} & ABCDE\\
 & 7568.899 & 4.283 & -0.773 & \cite{K07} & ACDE \\
 & 7583.787 & 3.018 & -1.885 & \cite{BWL} & E \\
 & 7586.018 & 4.313 & -0.458 & \cite{K07} & ACDE \\
 & 8108.320 & 2.728 & -3.898 & \cite{K07} & ABCDE\\
 & 8248.129 & 4.371 & -0.887 & \cite{K07} & D\\
 & 8471.743 & 4.956 & -1.037 & \cite{K07} & D\\
 & 8515.108 & 3.018 & -2.073 &  & D\\
 & 8616.280 & 4.913 & -0.655 & \cite{K07} & D\\
 & 8621.601 & 2.949 & -2.320 &  & D\\
 & 8698.706 & 2.990 & -3.452 & \cite{K07} & ABCDE\\
 & 8699.454 & 4.955 & -0.380 & \cite{NS} & ABCDE\\
 & 8710.404 & 5.742 & -5.156 & \cite{K07} & ABCDE\\
 & 8729.144 & 3.415 & -2.871 & \cite{K07} & ABCDE\\
Y I & 6402.006 & 0.066 & -1.849 & \cite{K07} & CD\\
 & 6435.004 & 0.066 & -0.820 & \cite{HL} & C\\
 & 6557.371 & 0.000 & -2.290 & \cite{K07} & CD\\
 & 6793.703 & 0.066 & -1.601 & \cite{K07} & CD\\
 & 8800.588 & 0.000 & -2.240 & \cite{CB} & ABCD\\
Y II & 7881.881 & 1.839 & -0.570 & \cite{Nil} & ACD \\
 Zr I & 7819.374 & 1.822 & -0.380 & \cite{Zrlines} & ABCDE\\
 & 7849.365 & 0.687 & -1.300 & \cite{Zrlines} & ABCDE\\
Nb I & 5189.186 & 0.130 & -1.394 & \cite{DLa} & ABCDE \\
 & 5271.524 & 0.142 & -1.240 & \cite{DLa} & AE\\
 & 5350.722 & 0.267 & -0.862 & \cite{DLa} & CDE\\
Ba I & 7488.077 & 1.190 & -0.230 & \cite{MW} & ABCDE\\
Ce II & 7580.913 & 0.327 & -2.120 & & CE\\ 
 & 8025.571 & 0.000 & -1.420 & \cite{MC} & CDE \\
 & 8404.133 & 0.704 & -1.670 & & C\\ 
 & 8716.659 & 0.122 & -1.980 & \cite{MC} & ABCDE\\
 & 8772.135 & 0.357 & -1.260 & \cite{PQWB} & ABCDE \\
Pr II & 5322.772 & 0.483 & -0.141 & & CD\\
Nd II & 5276.869 & 0.859 & -0.440 & \cite{MC} & CD\\
 & 5293.160 & 0.823 & 0.100 & \cite{HLSC} & CD \\
 & 5319.810 & 0.550 & -0.140 & \cite{HLSC} & CD\\
 & 5385.888 & 0.742 & -0.860 & \cite{MC} & CD \\
 & 5431.516 & 1.121 & -0.470 & & D \\
 & 7513.736 & 0.933 & -1.241 & \cite{MC} & CD \\
Sm II & 7042.206 & 1.076 & -0.760 &  & D \\
Eu II & 6437.640 & 1.320  & -1.998  & & D \\ 
 & 6645.061 & 1.380 & -0.516 &  & D\\ 
\hline

\end{tabular}

\end{table*}

\end{appendix}


\end{document}